\documentclass[12pt]{article}
\usepackage{epsfig}
\begin{document}

\begin{center}
\vspace*{6cm}

{\Large The Summary for Optimization of the Annular Coupled Structure}

\bigskip

{\Large Accelerating Module Physical Design}

\bigskip

{\Large for High Intensity Hadron Linac}

\vspace*{6cm}
{\Large V.V. Paramonov $*$}

\vspace*{3cm}

{\large KEK, Tsukuba, 2013}

\vspace*{2cm}

\end{center}

$*$ - permanent address - Institute for Nuclear Research of the RAS, 117312, Moscow, Russia
\newpage
\begin{abstract}
The normal conducting Annular Coupled Structure (ACS) is applied for 190-400 MeV 
part of high intensity proton linac for the J-PARC. The ACS operating frequency is 972 MHz. The J-PARC ACS is 
 strongly based on the results of previous investigations, especially results of Japan Hadron Project (JHP) research 
program in KEK. However, the design was revised and optimized to meet the 
requirements of reliability, operation efficiency and cost reduction.\\
The cells shape of accelerating cells was optimized in total energy range to have high shunt impedance value together with 
the careful matching with the decreased coupling cells. The design of the bridge coupler cells was optimized to simplify mass 
production and shape of RF input cell together with matching window were optimized for higher operational reliability.\\
Collected and adjusted all together, these 
modifications result in the significant effect. 
The ACS module design doesn't lose to another possible accelerating structures in RF parameters 
and dimensions. Previously developed for L-band ACS fabrication technique is applied to $972$ MHz ACS construction.
The realized solutions have the reserve for ACS module use with very high heat loading during operations with $15\%$ duty factor,
the highest heat loading at present time, as compared to accelerating modules with another normal conducting accelerating structures,
realized in hadron linacs.\\
This report contains the descriptions and explanations of changes in the ACS module physical design for the  J-PARC linac.  
\end{abstract}
\newpage
\tableofcontents
\newpage
\section{Introduction}
In the second part of the 20-th century the development of high intensity normal conducting linacs with the final 
energy of hydrogen ions up to (600 - 1000) MeV with the average beam current up to (0.5 - 1.0) mA were under development 
in several countries. The first such linac was constructed in Los Alamos, USA, at the energy 800 MeV, pulse beam current 17 mA and 
average beam current 1.2 mA, starting with the first beam  in 1972. The second such linac, with the designed energy 600 MeV, pulse 
current 50 mA and average current 0.5 ma, was constructed in USSR, INR of the RAS, starting with the first beam in 1990 and 
now operating, unfortunately, not with full power.\\
\begin{figure}[htb]
\centering
\epsfig{file=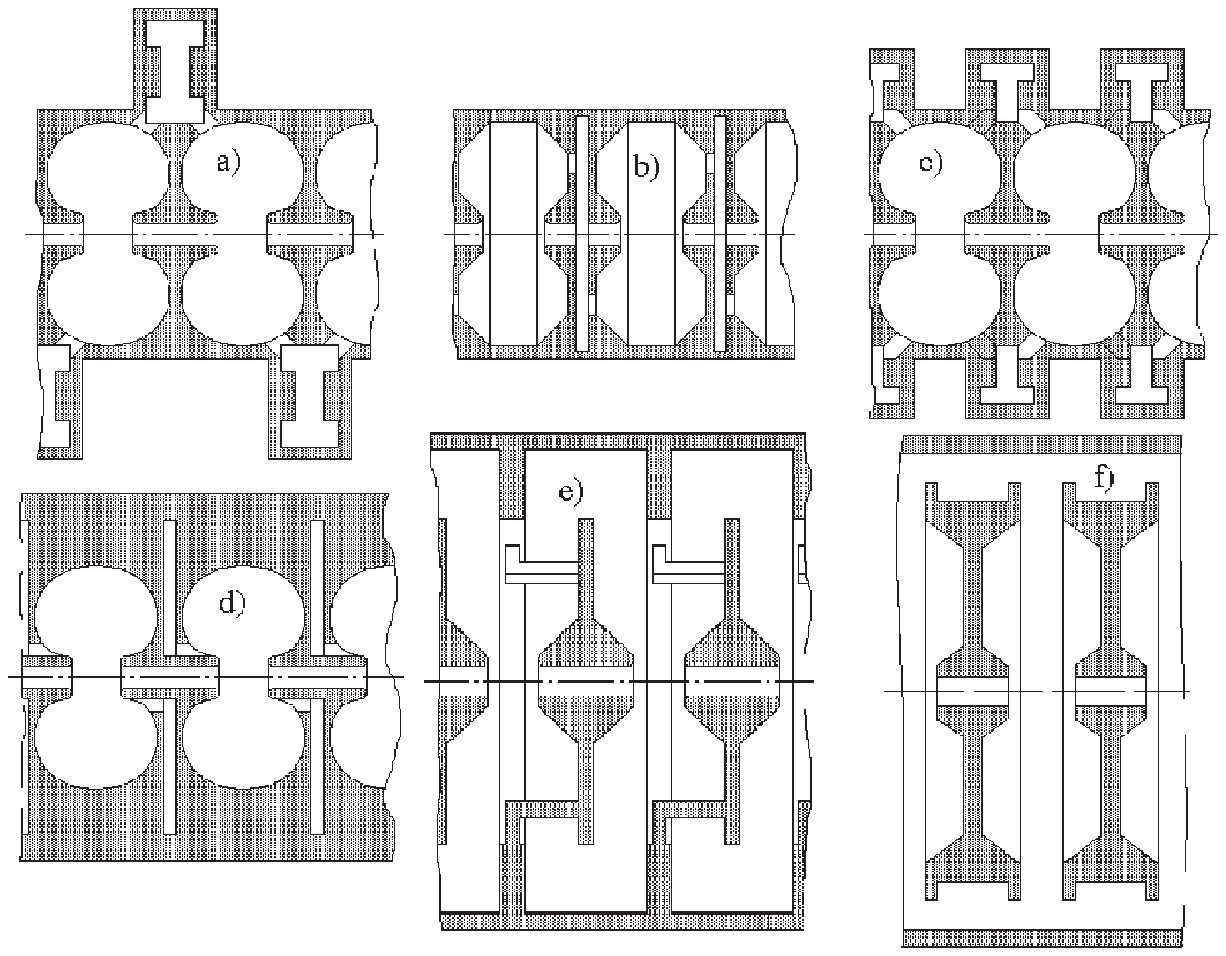, width =115.0mm}
\begin{center}
 Figure 1: Schematic sketches for coupled cells structures for high particle velocity. a) - Side Coupled Structure (SCS), 
b) - On axis Coupled Structure (OCS), c) Annular Coupled Structure (ACS), d) - structure with coaxial coupling cells, e) Disk And 
Washer (DAW) structure, f) structure with shaped washers.
\end{center}
\label{1f}
\end{figure}
During development of such linacs, named that time as 'meson factories', two important statement for normal conducting 
accelerating structures in these linacs, were formulated.\\
There are no single accelerating structure, which can overlap the total energy range of such hadron linac.\\
For high intensity linacs so called compensated accelerating structures should be used, which combine the high RF efficiency 
with the high stability of accelerating field distribution with respect to errors in manufacturing and tuning, beam loading. 
The compensated are named structures, in which at operating frequency coincide frequencies of two modes - accelerating mode 
and coupling mode, with conjugated parity of field distributions, \cite{dome}. Another definitions of these structures, used 
in the literature, are bi-periodical structures or coupled cell structures.\\ 
For the initial part of the linacs with the energy up to (50 - 80) MeV the Drift Tube Linac (DTL), or Alvarez structure, is used with 
post couplers (for field distribution stabilization) until now without principal changes. In the energy range from $80$ MeV to 
$~ 150$ MeV based on DTL structures, Separated DTL (SDTL) or Coupled Cell DTL (CCDTL) can be applied.\\ 
Anyhow, with the beam energy increasing RF efficiency of DTL-based structures decreases and another accelerating structure should be applied.
There were a lot of proposals of such structures for the high energy part of the linac. Most typical are shown schematically in the 
Fig. 1.\\ 
Accelerating structure for the high intensity hadron linac should satisfy to the set of different requirements, which are some times 
contradictory - required RF parameters, operational reliability and stability, vacuum conductivity, technological aspect, including 
the price of construction. It stimulates research and development for new structures. But very important, decisive points are the  
level of development and demonstration of the required parameters of the full scale structure at high RF power. It confirms the possibility 
to create the accelerating system with required parameters.
\section{Proven accelerating structures for high energy part of the linac}
\begin{figure}[htb]
\centering
\epsfig{file=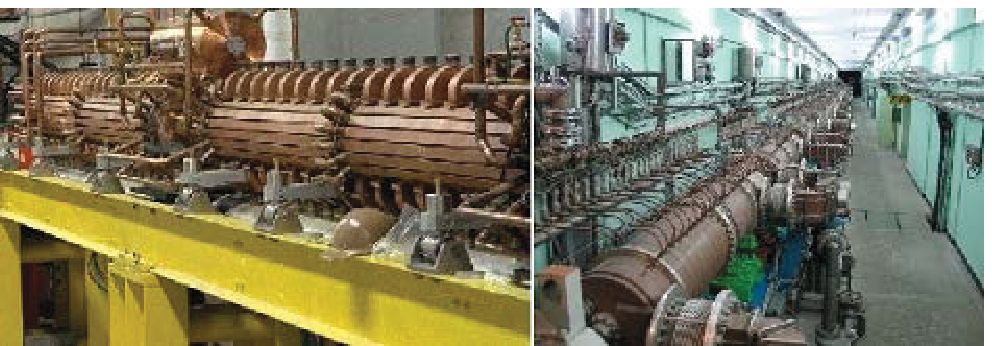, width =150.0mm}
\begin{center}
 Figure 2: The SCS structure in the high energy part of the LANSCE linac (left) and the high energy part of the INR linac with DAW 
structure (right).
\end{center}
\label{2f}
\end{figure}
The first accelerating structure, realized in the high energy part of the linac, Fig. 2, SCS, invented and investigated in LANL, 
\cite{knapp}, \cite{knappc}. Later this structure was used in FNAL injector linac energy upgrade to 400 MeV and also applied 
in SNS linac in the room temperature part. The schematic view of this structure is shown in Fig. 3b. The structure has a coupling 
coefficient value $k_c \approx (4-5)\%$. Due to position of coupling slots oppositely in diameter, in the field distribution 
of accelerating cells there is dipole addition and at the beam axis there is not zero transverse field. Due opposite direction 
in adjacent cells the effect of transverse field for the synchronous particle cancels in the first order in $\frac{\delta \beta}{\beta}$, 
but the second order effects remain. Here $\delta \beta$ is the relative increasing of the particle velocity at one accelerating gap.\\
\begin{figure}[htb]
\centering
\epsfig{file=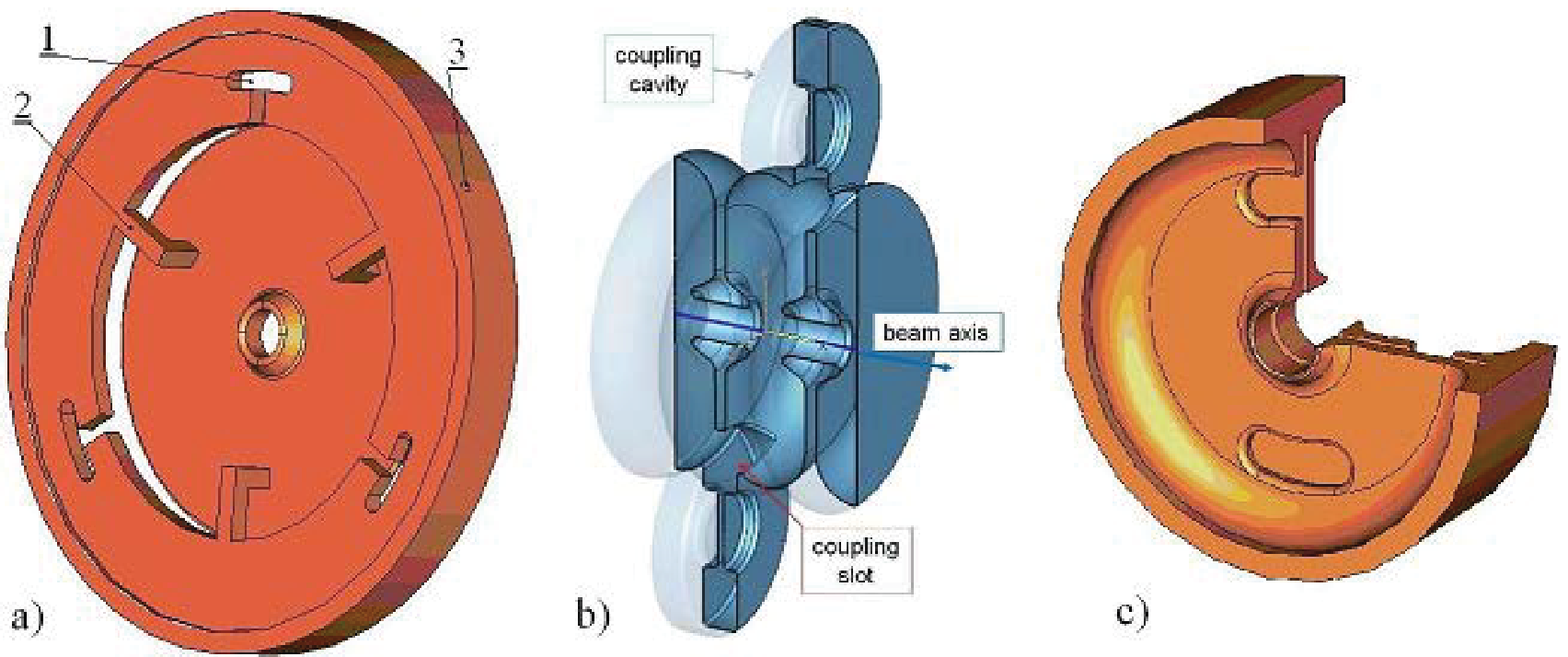, width =150.0mm}
\begin{center}
 Figure 3: Accelerating structures, realized in linacs. a) DAW for INR linac, 1 - slots for HOM displacement, 2 - support, 3 - input/
output for cooling water; b) SCS, c) OCS - electron linacs.  
\end{center}
\label{3f}
\end{figure}
In the USSR development program  of 'meson facility' linac ACS and DAW were invented, \cite{ACS_inv}, \cite{DAW_inv}. Both structures 
were tested at high RF power level, \cite{murin}, and, due to higher coupling coefficient $k_c \approx (40-50) \%$, the DAW structure, 
Fig. 3a, was chosen as realized in INR linac, Fig. 2. Together with the high $k_c$ value, DAW has an excellent vacuum conductivity. 
The washers are connected to disks with L-type supports, in which the cooling channels for water supply in the washer are placed.    
The structure has a severe High Order Modes (HOM) problem, and HOM's are placed below, in the vicinity, and above operating frequency. 
In the structure design the special selective resonant elements - slots (3, in Fig. 3a) were introduced. These slots do not interact 
with accelerating mode and do not deteriorate operating RF parameters. But, interacting with HOM's, slots result in the displacement 
of HOM's from the vicinity of operating frequency. Another solution for HOM's displacement was developed later with bi-periodic L-type 
supports by Y. Iwashita. For this structure there are at least two solutions for RF HOM problem. For the range of particles velocity 
$\beta = (0.4 \div 0.8)$ from $95\%$ to $80\%$ of the total RF losses are in the washers. For application in linac with 
the high duty factor value the heat load exceeds in order the value, proven in INR linac. Because water supply can be 
realized only through thin L-type supports, this problem looks difficult.\\
The On axis Coupled Structure, Fig. 3c, is widely used in compact electron linacs. For protons acceleration this structure was considered
too, but rejected, due essential reduction of effective shunt impedance $Z_e$ for medium $\beta$ range and possibility of a stable 
multipactoring in coupling cells.
\begin{figure}[htb]
\centering
\epsfig{file=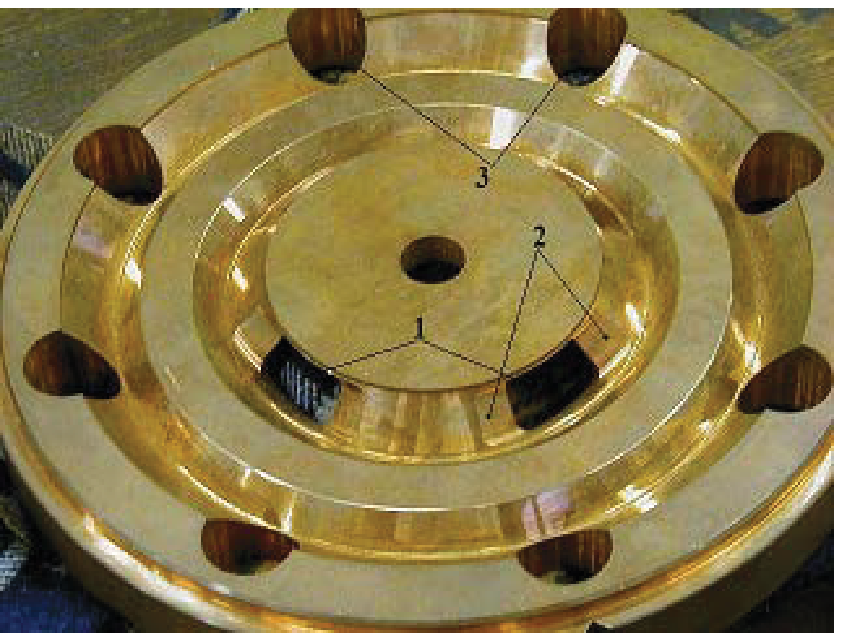, width =90.0mm}
\begin{center}
 Figure 4: JHP ACS unit shown here from the side of coupling cell. 1) - coupling slots, 2) - tapering, 3)- vacuum ports.
\end{center}
\label{4f}
\end{figure} 
\subsection{The ACS improvements during JHP program} 
During JHP research and development program in KEK \cite{acskek}, under leadership of Y. Yamazaki, the ACS design for L-band frequency range, Fig. 4, was essentially, in 
some parameters qualitatively, improved. The number of slot was increased from two to four. It ensures decoupling of accelerating mode 
with $TM_{110}$ HOM mode, 
which exists in coupling cells not so far from operating frequency. The tapering, applied near coupling slots, (2 in Fig. 4) allows 
$k_c$ value $\approx 5\%$ together with significant thickness of the web between accelerating and coupling cells. The sufficient 
web thickness allows a placement of cooling channels. Elaborated and effective cooling circuit was developed, \cite{acsrefco}, 
ensuring stable ACS operation with high heat loading. The fabrication technique for L-band ACS was developed and tested also,
\cite{acsrefmf}. The multi-cell bridge cavity, \cite{acsbrid}, was developed to overlap the space required for focusing elements.  
In the bridge cavity are placed fast movable tuners for operating frequency fast control. The cooling water is used only for 
heat removal. The entire cooling system becomes much more simple, as compared to the solution of the frequency control by 
change of cooling water temperature or flow.\\
Totally, in JHP research was developed the concept of accelerating module, Fig. 5. 
Finally, L-band ACS module was successfully tested at the high level of RF power, \cite{acsrefrf}.\\  
\begin{figure}[htb]
\centering
\epsfig{file=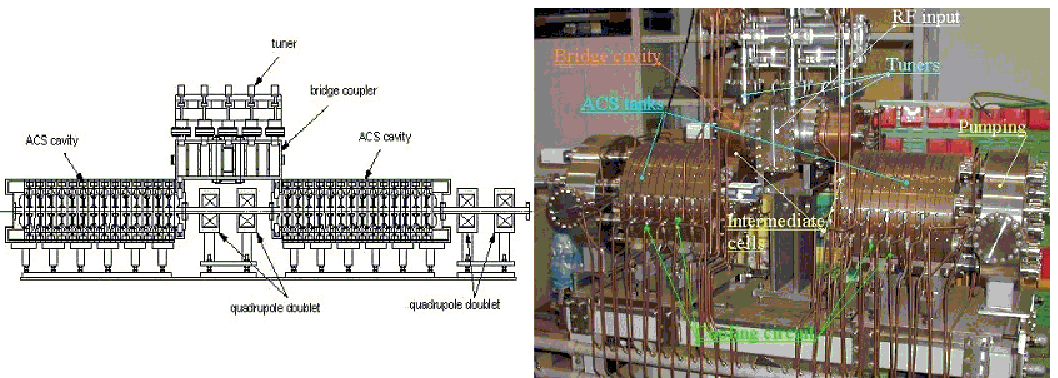, width =120.0mm}
\begin{center}
 Figure 5: Schematic sketch of the ACS accelerating module (left) and the L- band ACS module for high RF power test in JHP program.
\end{center}
\label{5f}
\end{figure}
Basing on:
- results of comparative estimations for different accelerating structures for high intensity hadron linac;\\
- proven results of the L-band ACS structure performance and the developed fabrication technique;\\
- accumulated experience for L-band ACS module construction with required parameters;\\
- proven results of L-band accelerating ACS module and promising expectations and estimations for operating frequency 972 MHz;\\
the development of accelerating module with ACS structure was chosen as the base line for normal conducting module for J-PARC linac.
\section{Optimization of ACS cells. Design parameters.}
The direct scaling from the L-band ACS design is not a solution, because leads to increasing of transverse dimensions in 1.33 times and 
structure weight per unit length in 1.75 times. In this way the developed brazing technique, applied also for SCS in LANL and DAW in INR,
meets strong problems in the uniformity and reliability of brazed joint. After SCS construction in LANL with $f_{op}=805 MHz$, 
where $f_{op}$ is the operating frequency, there was conclusion, 
that the possibilities of brazing technique limit possible SCS operating frequency to $f_{op} > (600 - 700) MHz$. Later, the problems 
in the construction of SCS module with $f_{op} =704 MHz$ for high RF power test were not the last reasons to change the structure.\\
In ACS optimization transverse dimensions should be reduced essentially. It are defined in ACS by diameter of coupling cells. 
But another points - vacuum ports and connection with accelerating cell - are also important.\\
During development the achieved L-band ACS performance should not be lost. Also the high heat capability should be saved - ACS module 
should have a possibility for operation with $15\%$ duty factor.
\subsection{Accelerating cells optimization}
The procedure and technique of cells optimization is described in \cite{acsparam}. The final choice was done after several 
iterations in the mutual adjustment of coupling and accelerating slots and here mainly reasons, results and conclusions are given.\\
For the wide range optimization, described in \cite{acsparam} two shapes of accelerating cell, shown in Fig. 6b and Fig. 6c, were 
used.\\ 
\begin{figure}[htb]
\centering
\epsfig{file=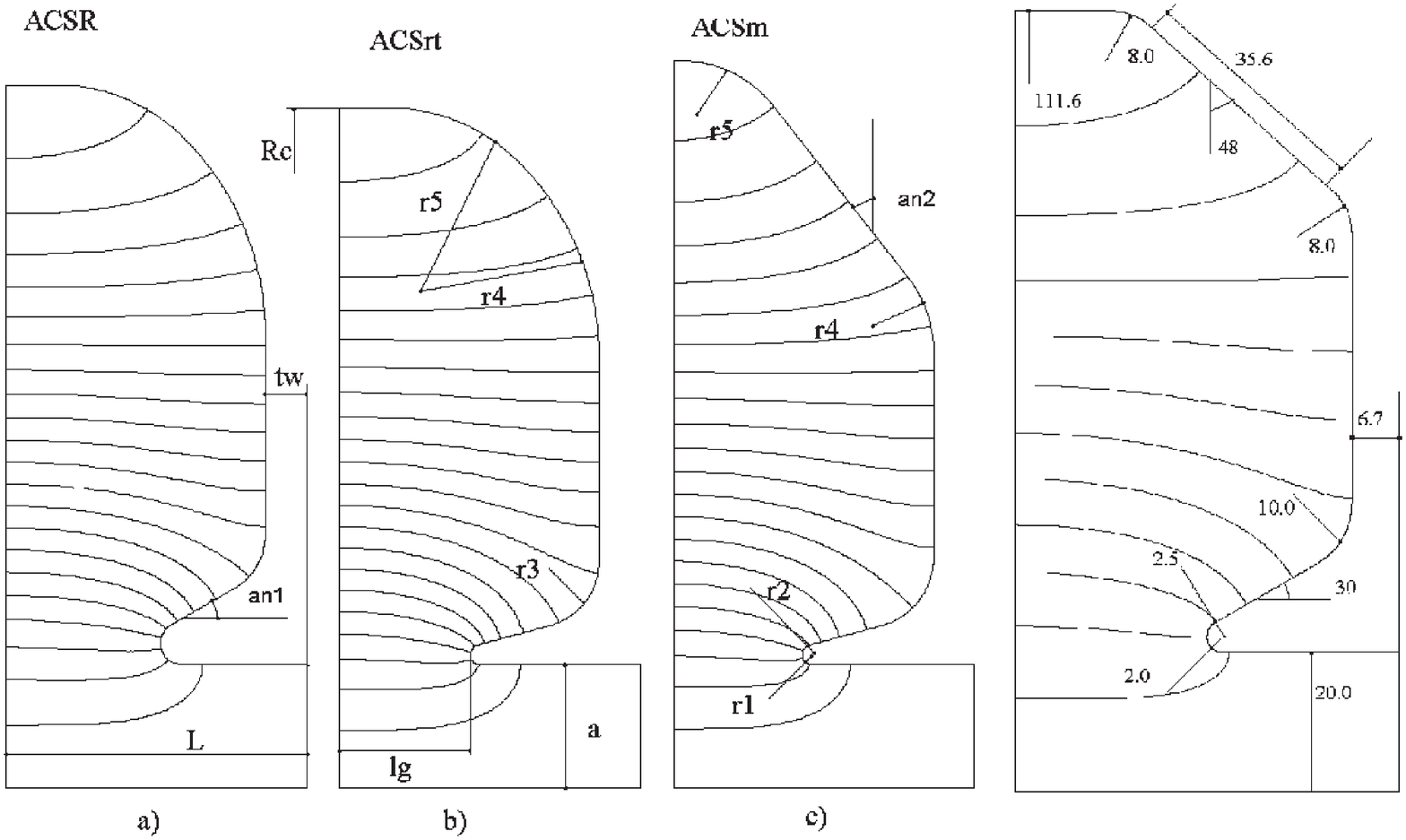, width =125.0mm}
\begin{center}
 Figure 6: Considered shapes of the ACS accelerating cell a) - the scaled L-band shape, b) - the shape with two outer rounding, c) -
the shape with a conical part, d) - the final choice. 
\end{center}
\label{6f}
\end{figure}  
For reference, the shape of L-band ACS was scaled to $f_{op}=972 MHz$, Fig. 6a. Two another shapes differ in outer part. The shape in 
Fig. 6b has two radii of outer rounding. The shape in Fig. 6c has the conical part with small adjacent rounding, both 
for quality factor small improvement and to provide a smooth surface. The goal of this modifications is to find the most convenient 
and effective solution for accelerating cell outer shape for connection with coupling cell.\\
The drift tube nose has two radii of rounding, $r_1$ and $r_2$ in Fig. 6c. The maximal electric field at the surface $E_s$, which determines 
an electrical strength of the structure and possibility of breakdowns, takes place at the upper nose part. As the result of applied 
wide range study, we can have clear representation about the influence of each geometrical parameter on structure RF parameters. 
In Fig. 7a is shown the surface of the effective shunt impedance $Z_e$ on two variables - the radius of lower rounding $r_1$ and 
the gap ratio $\alpha$ for another free parameters being fixed. One can clearly see significant $Z_e$ reduction with $r_1$ increasing, 
while $r_1$ increasing affects $E_s$ slightly. We can all time find the relation $r_1$ and $r_2$ to improve $Z_e$ value but to keep 
$E_s$ below required value.\\      
\begin{figure}[htb]
\centering
\epsfig{file=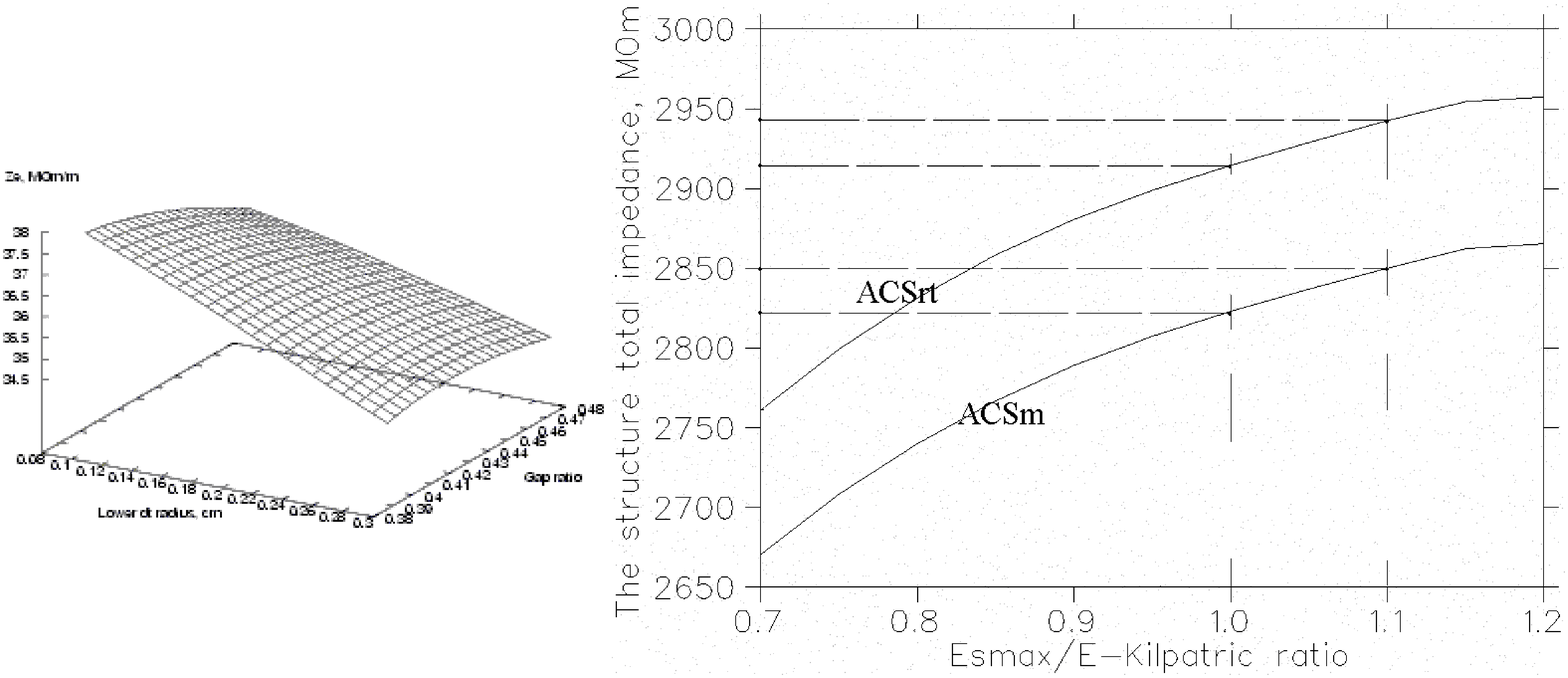, width =125.0mm}
\begin{center}
 Figure 7: The surface $Z_e(r_1,\alpha)$ to illustrate $r_1$ influence (a) and plots of of the total $Z_e$ value for the total ACS 
system in dependence on $E_s/E_k$ ratio (b).
\end{center}
\label{7f}
\end{figure}
For both shapes under consideration the influence of cell dimensions on RF parameters was considered assuming the total ACS 
structure for energy range from 190 MeV to 400 MeV, $\beta =(0.556 \div 0.713)$. Some dimensions were fixed, considering ACS RF 
parameters in relationships with beam dynamics and operational reliability requirements.\\
Instead of essential $Z_e$ increasing with decreasing the aperture radius $a$, this radius was fixed to $a=20 mm$ for the total 
ACS system, taking into account beam dynamics conditions, especially taking care for matching with previous linac part.\\
 The total ACS $Z_e$ value also can be increased by $t_w$ reduction - the web thickness between adjacent accelerating cells. But we have to place 
inside web effective cooling channels and foresee the sufficient thickness of material between cooling cannels and vacuum volume of 
accelerating cell, taking into account effect of high temperature brazing. The web thickness was fixed to $2t_w=13.7 mm$ for the entire 
ACS system.\\
Instead of cooling channels are placed close to the drift tube as possible, the heat evacuation from the drift tube nose, the point 
with the maximal $E_s$ value, is due to natural copper heat conductivity only. The drift tube cone angle was fixed to $30^o$ as the 
compromise between possible $Z_e$ increasing by angle reduction and heat evacuation conditions without cooling channels inside the 
drift tube.\\
In Fig. 7b are shown plot of $Z_e$ value for the total ACS system in dependence on $E_s/E_k$ ratio, where $E_k$ is the Kilpatric 
threshold value for $f_{op}=972 MHz$. As one can see, plot show saturation with $E_s/E_k$ increasing. Increasing $E_s/E_k$ from 
1.0 to 1.1, we can expect $Z_e$ increasing at $25 MOm$, or relative increasing $\frac{\Delta Z_e}{Z_e} < 1\%$. With 
increasing of time for conditioning and risk of possible breakdowns during operation we have no adequate compensation in RF 
efficiency. For the total ACS system the maximal surface electric field was fixed as $E_s \sim 1.0 E_k$.\\
For another dimension of accelerating cell the RF parameters of the entire ACS system were considered in the option "all variable" 
to get the maximal $Z_e$ value with limitations for $a, t_w$ and $E_s$, applied before. After that the fixed parameters 
for entire ACS system were founded to have the minimal deviation from the optimal $Z_e$ value. The relative difference was found as 
$1.2\%$ and for cost reduction in ACS production the dimensions of accelerating cells were fixed, as shown in Fig. 6d. The adjustment 
of operating frequency for each $\beta$ is by adjustment of gap length between drift tubes.\\
As one can see from the plots in Fig. 7b, the price for conical part in the accelerating cell shape is $Z_e$ reduction at $\sim 3\%$ 
as compared to the totally rounded shape. But it is not the final result for the ACS J-PARC structure - it is the intermediate result 
in 2D, approximation without connection with coupling cell. We have to get the best final result in 3D case.
\subsection{Coupling cells optimization}
To decrease ACS transverse dimension, the "radial" cell orientation, realized in L-band ACS, Fig. 8a, was changed to the "longitudinal" one, 
Fig. 8b.\\        
\begin{figure}[htb]
\centering
\epsfig{file=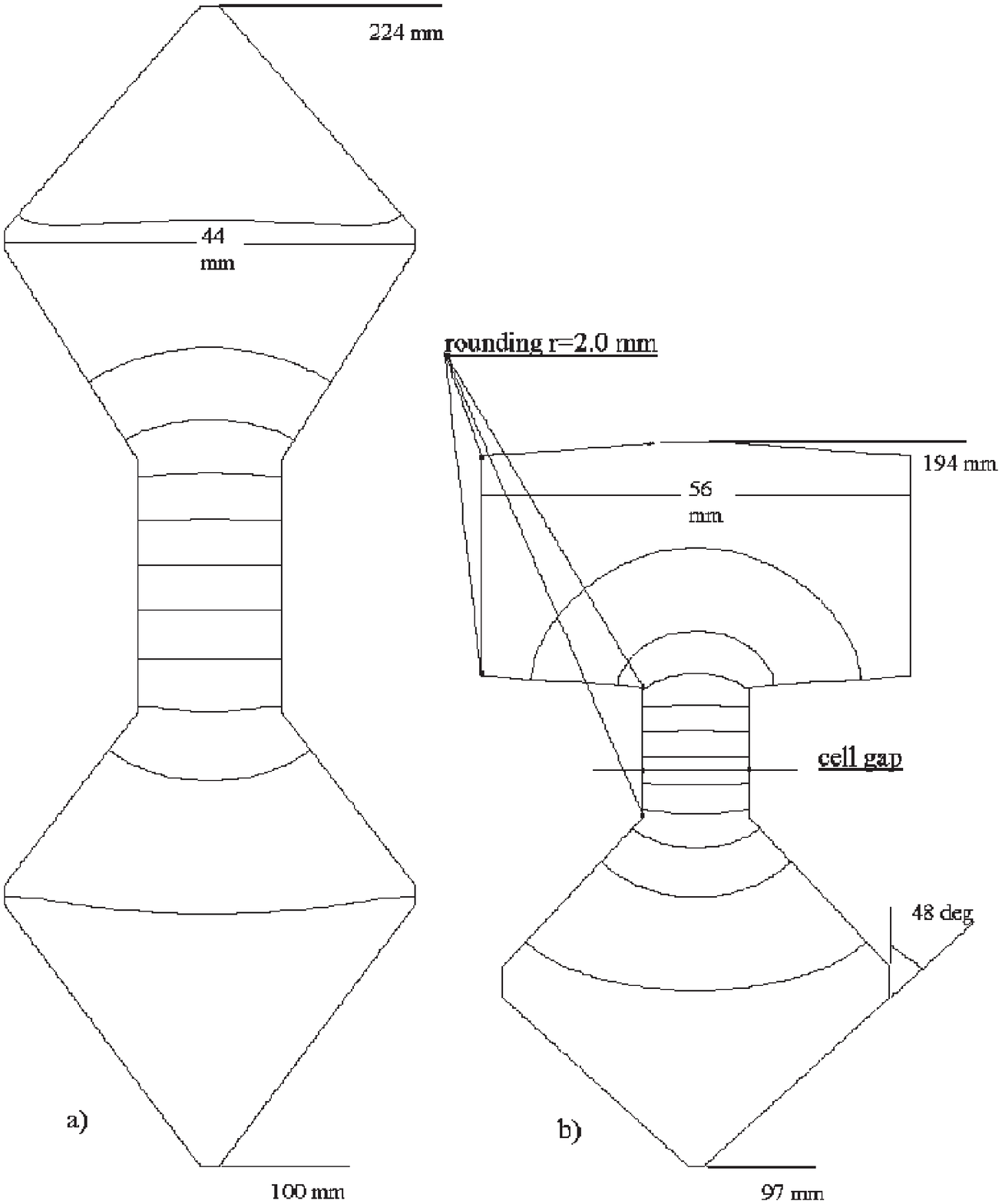, width =95.0mm}
\begin{center}
 Figure 8: The coupling cell shapes for ACS structure, a) the scaled from the L-band ACS, b) - the modified shape.
\end{center}
\label{8f}
\end{figure}
The shape and dimensions for the lower part of the cell were chosen for better adjustment with accelerating cell, comfortable 
tool access in surface treatment and for higher, than in upper part, magnetic field density for higher coupling coefficient. 
The gap length in the cell, Fig. 8b, was reduced slightly, to keep conditions for multipactor discharge in coupling cells as ensured 
"not possible", \cite{mult}, and do not increase the sensitivity of coupling cell frequency to deviation of dap length 
during mechanical treatment and brazing.\\
The vacuum ports are shifted inside the cell, crossing the outer cell part and the cell gap region too. The cells with 
the modified shape were analyzed for breakdown possibility with cells excitation due to accelerating cells detuning and during 
transient and the big reserve in the electric strength was detected.\\
For mass production simplification the dimensions of coupling cells are the same through the total ACS system. Later the cell shape 
was corrected slightly for further simplification - the outer cell part was modified to flat.\\
In such design the transverse ACS dimensions are defined by the outer coupling cell radius.      
\subsection{Cells adjustment and coupling slots orientation}
The value of coupling coefficient $k_c$ between accelerating and coupling cell depends on the dimensions, position and mutual orientation 
of coupling slots. For the structures, coupled with slots for one slot 
\begin{equation}
k_c \sim \frac{ h_s l_s^3 H_a H_c}{t_s \sqrt{W_a W_c}},
\label{1e}
\end{equation}
where $l_s$ and $h_s$ are the lengths and the height of the slot, $t_s$ is the web thickness between cells in the place of the slot, 
$H_a, H_c$ are the magnetic field at the slot for accelerating and coupling cells, $W_a, W_c$ are the stored energy in accelerating and 
coupling cells respectively. Instead of a strong dependence $k_c \sim l_s^3$, the slot length increasing leads to increasing 
of RF current density at the slot ends and $Z_e$ reduction. At first, the slot height $h_s$ was adjusted to maximal comfortable value 
and tapering angle was optimized. Finally, the slot dimensions were fixed to $h_s=27.6 mm$ and the slot opening angle (from the axis) 
$33^o$.\\ 
The slot edges from the side of accelerating cell were rounded with $r=4 mm$. As compared to sharp $90^o$ edges, it results, according 
simulations, in $k_c$ increasing at $\sim 0.3\%$, (from $5.58\%$ to $5.85\%$, $Q_a$ increasing is several percent's and thermal 
stress reduction in the slot region. (Later, in ACS optimization for mass production, slot rounding was changed to chamfer without 
degradation of another performances).\\   
If slots at the opposite sides of the cell are placed "face to face", the partial coupling cancellation takes place. 
The maximal value of coupling coefficient can be achieved when the slots at one side of the cell are placed in between of slots 
at opposite side of the cell. Such slots orientation was realized in the L-band ACS and the procedure was established to 
determine cells frequencies for RF tuning before brazing. Such configuration realize complicated symmetry of the structure - 
simultaneous translation and rotation, \cite{acskek}. Such structures have no planes of mirror symmetry, where shorting plates can 
be applied and approximated method of cells detuning was used for frequencies estimations.\\
The coupling cancellation effect depends on the longitudinal distance between slots. The coupling cell is short enough and 
"face to face" slots orientation at the opposite sides leads to strong $k_c$ reduction. In the coupling ACS cells slots at one side 
of the cell are shifted at $45^o$ with respect to slots at the opposite side. In SCS coupling cells "face to face" slots orientation,
Fig. 3b, is realized just from necessity and coupling cancellation should be compensated by slots increasing or stronger magnetic 
field at the slots.\\ 
Simulations have shown, \cite{acsparam}, that for the worth case of the short accelerating cell $\beta=0.556$ the difference in $k_c$ value is not 
so big - $k_c=6.25\%$ for rotated slot position and $k_c=6.05\%$ for opposite slots position, for example, with the relative difference 
$\frac{\Delta k_c}{k_c} \approx 3.3\%$. For the higher $\beta$ values this difference is even smaller. But with opposite slot position 
we get a plane of mirror symmetry in the middle of accelerating cell and can utilize much simpler procedure for cell frequency 
measurements during RF tuning. At the plane of mirror symmetry the field of accelerating mode satisfies to electric wall 
boundary conditions and shorting plate can be used. The measurement of accelerating cells frequency is evident, similar to 
shown in Fig. 9a or without detuning rods in pumping ports. For coupling mode frequencies measurement non direct method is suggested, 
\cite{acsparam}, because the coupling mode with magnetic wall boundary condition can not be excited. In the assembly with 
two ACS periods we measure frequencies for $0-$ and $\pi/2-$ mode of the chain from two coupling cells, because accelerating 
cell is detuned drastically. From the measured values we can extract the frequency for $\pi-$ mode, which corresponds to the 
coupling cell frequency.\\
\begin{figure}[htb]
\centering
\epsfig{file=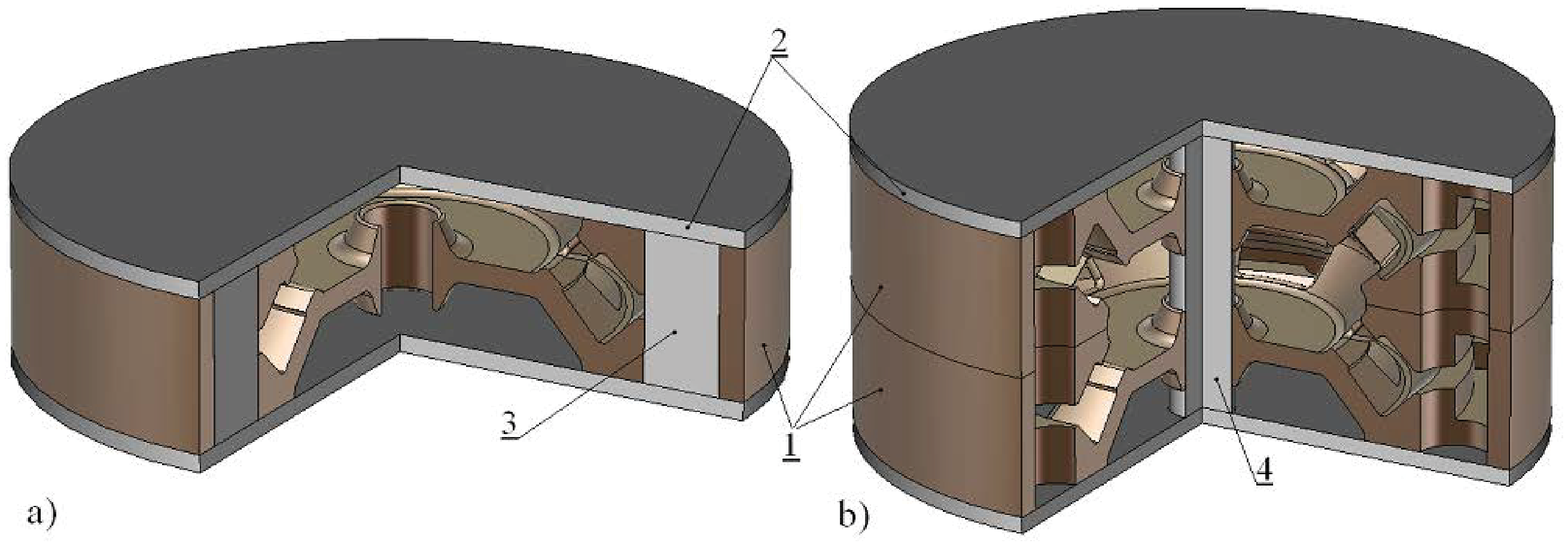, width =115.0mm}
\begin{center}
 Figure 9: Schematic sketches for frequency measurements of accelerating mode (a) and coupling mode (b) frequencies. 1) - ACS periods, 2) - 
shorting plates, 3) - detuning rods for coupling cell, 4) - detuning rod for accelerating mode. 
\end{center}
\label{9f}
\end{figure}
All time coupling slots deteriorate the axial symmetry of the cell, resulting in multipole additions in the field distribution. 
For four slots at each side of the cell this addition starts all time from quadrupole and higher. But for the rotated slot position 
the quadrupole addition has the even dependence of transverse field components with respect to symmetry plane, and for opposite slot 
position - the odd  dependence with much faster decay of additions from the slot to the structure axis. It is additional preference 
of coupled slot structures with mirror symmetry plane.\\
Basing on obtained results for not so big $k_c$ decreasing and practical preference of opposite slot position at opposite sides of 
accelerating cell, this configuration is chosen for J-PARC ACS.\\         
Normally in the slot coupled structures the coupling coefficient value can be achieved at the expense of $Z_e$ reduction with the 
rate $\sim (1-2)\%$ reduction in $Z_e$ for $1\% k_c$ increasing. For the slot coupled structures with the rounded outer part of accelerating cell 
the adjustment with coupling cell is not convenient. One can see the complicated surface in the connecting region for SCS, Fig. 3b.
During SCS cells production for SNS linac additional mechanical treatment was required to remove sharp edges and provide tolerable slot shape, 
\cite{scssns}.\\
\begin{figure}[htb]
\centering
\epsfig{file=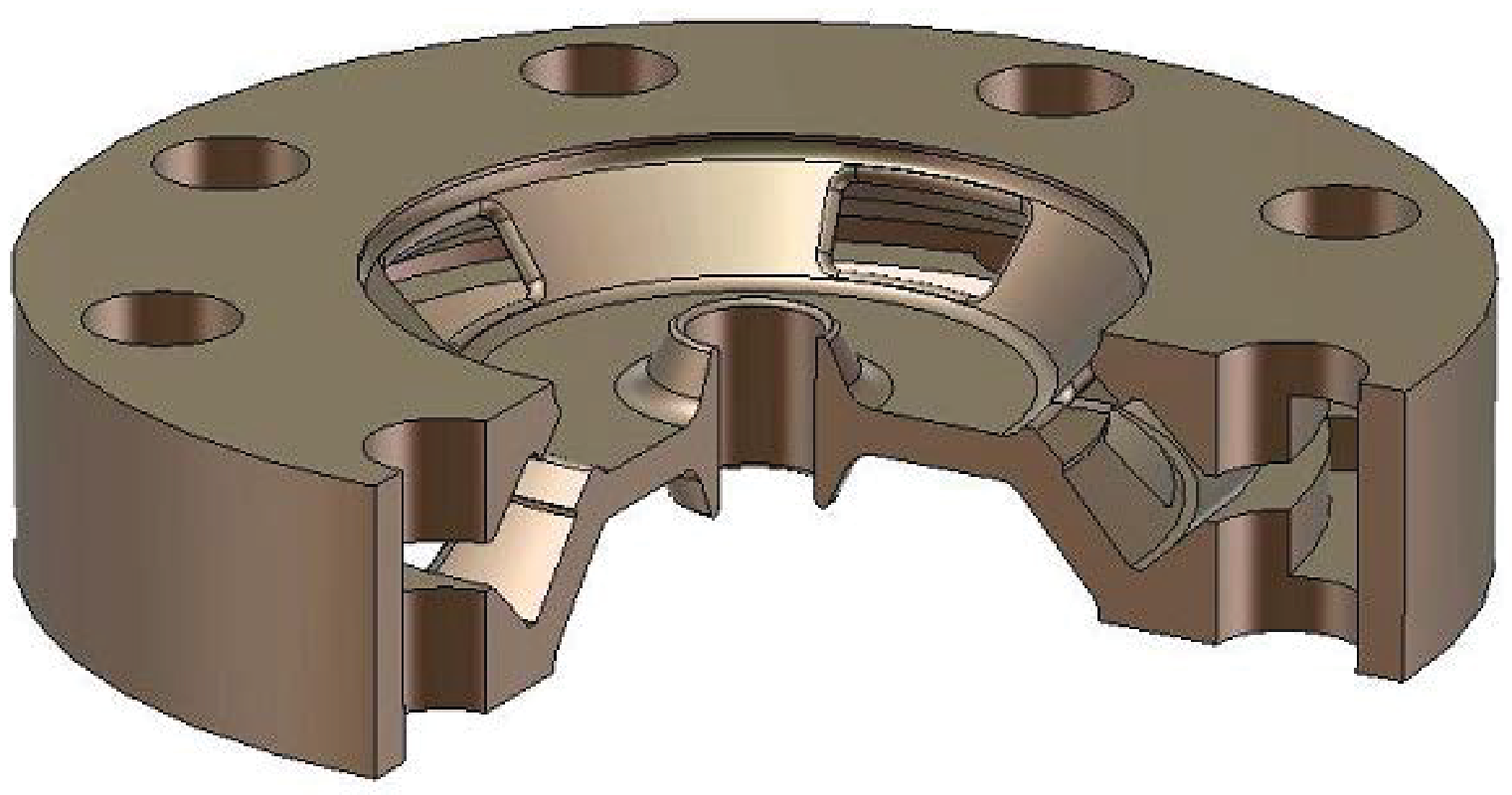, width =115.0mm}
\begin{center}
 Figure 10: The final configuration of the J-PARC ACS period.
\end{center}
\label{10f}
\end{figure}
With the conical part in the shape of accelerating cell, Fig. 6c, we get a strong mechanical design, with controllable slot shape 
and reduced slot length. For the shorter slot the density of RF current near slot ends is lower and in the total 3D structure we 
recover the reduced at $3.3\%$ in 2D case $Z_e$ value. Comparative 3D simulations have shown, that for the same 
$k_c$ value and the same modified coupling cell the structures with rounded accelerating cell, Fig. 6b, and the structure with 
conical part in accelerating cell, Fig. 6c, differ in $Z_e$ at $ < 1\%$.\\
\begin{table}[htb]   
\begin{center}
\centering{Table 1: The calculated parameters of ACS cells for 972 MHz.}
\begin{tabular}{|l|c|c|c|c|c|c|}
\hline
 $\beta$  &$k_c, \% $ & $ Q_a$ &$Z_e, \frac{MOm}{m}$, & $T$ & $\frac{Z_e}{Q_a}, \frac{kOm}{m} $ & $\frac{E_{smax}}{E_oT}$\\
\hline         
 0.5583   & 5.86      & 16190  & 28.1                 & 0.8366  &  1.735                        & 7.385         \\
\hline
 0.6427   & 5.58      & 18720  & 33.64                & 0.8313  &  1.797                        & 7.691          \\
\hline
 0.7085   & 5.32      & 20690  & 37.64                & 0.82574 &  1.820                        & 7.381         \\
\hline
\end{tabular}
\label{1t}
\end{center}
\end{table} 
In ACS $k_c$ depends also on the thickness of the web between coupling and accelerating cells. For web thickness $12 mm$ $k_c 
\approx 7.2\%$ is possible. But in this web should be placed cooling channels to supply fluid for cooling of the accelerating cell.
The web thickness is chosen $16 mm$, which is comfortable and reliable for placing of cooling channels $7 mm$ in diameter.\\
With modification of mutual slot positions in accelerating cells the cooling circuit is modified slightly in the directions 
of cooling fluid, \cite {acsparam}, but without decreasing of cooling ability.\\ 
Developed with all modifications, described above, the configuration of ACS for J-PARC linac is shown in Fig. 10.\\
Calculated in 3D approximation RF parameters, \cite{acsparam} are illustrated in the Table 1, where $Q_a$ is the quality 
factor for accelerating mode, $T$ is the transit time factor value. For 2D approximation more result one can find in \cite{acsparam}. 
\subsection{ACS with the large aperture radius}
For beam parameters matching between the linac and RGS special cavities - debunchers - with essentially increased aperture radius $a$ are 
required. The general study of different accelerating structures for strongly increased aperture has been performed and described 
in \cite{bighole}. Results shown, that $Z_e$ reduction with $a$ increasing takes place in different structures with the same scale. 
In ACS structure the deterioration of dispersion curve due to increasing of neighbor coupling between accelerating cells, starts after 
$2a > 100 mm$. No reasons were found to use for debancher another structure. After mutual consideration of contradictory requirements 
of beam dynamics and ACS RF parameters, the aperture diameter was fixed to $2a=70 mm $ for Debuncher 1 and $2a=85 mm$ for Debuncher 2. 
The maximal possible value of $Z_e$ is required all time at least to relax RF power requirement.\\
\begin{figure}[htb]
\centering
\epsfig{file=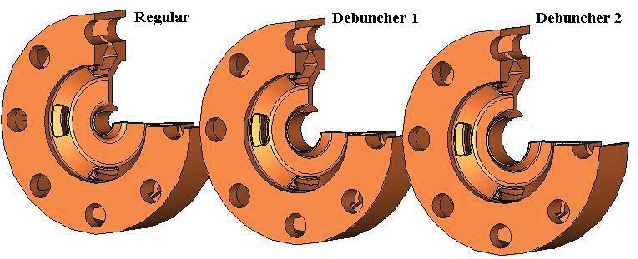, width =135.0mm}
\begin{center}
 Figure 11: ACS structure, regular cells, $2a=40 mm$ (a), cell for Debuncher 1, $2a=70 mm$ (b) and cell for Debuncher 2 (2a=85 mm) (c).  
\end{center}
\label{11f}
\end{figure}
For the highest $Z_e$ value the optimal value of accelerating gap ratio $\alpha$ depends on $\beta$ and $a$ values. For $\beta=0.7131$ 
the dependence $Z_e(\alpha)$ was studied for accelerating cells with increased aperture first in 2D approximation, but at the frequency $995 MHz$, 
taking into account future frequency decreasing by coupling slots. After that several options were considered in 3D simulations.\\
To realize the optimal $\alpha$ value, we have to adjust accelerating cell radius $R_c$. But the dependence $Z_e(\alpha)$ is 
smooth near point of maximum and some deviation from optimal point are possible. For Debuncher 1 application of the same cell radius, 
as for regular ACS cell, $R_c=111.6 mm$, results in relative $Z_e$ reduction only at $2\%$ and is not significant. For this cavity 
the outer part, including coupling cell, should be the same as for regular ACS.\\
For Debuncher 2 realization of $r_c=111.6 mm$ leads to relative $Z_e$ reduction at $6.3\%$ with respect to optimal value. But optimal 
$\alpha$ value results in $R_c=119.6 mm$ and also requires rearrangement of coupling cell dimensions ($R_cc$) and increasing of outer diameter. 
It is not a problem in simulations, but can lead to additional set of axillary tools in cavity construction.\\
The schematic pictures for ACS cells for debunching cavities are shown in Fig. 11 in comparison with regular cell. The calculated results 
for debunching cavities are listed in the Table 3. After consideration of the total set of sequences, the Debuncher 2(a) option was 
selected for realization.
\begin{table}[htb]   
\begin{center}
\centering{Table 2: The calculated parameters of debunching cavities.}
\begin{tabular}{|l|c|c|c|c|c|c|}
\hline
 Parameter & Debuncer 1 & Debuncer 2(a) & Debuncher 2(b)  \\
\hline         
 $\beta$               & 0.7131         & 0.7131      & 0.7131        \\
 $2a, mm$              & 70             & 85          & 85            \\
 $\alpha$              & 0.5214         & 0.5025      & 0.6306        \\
 $R_c, mm$             & 111.6          & 111.6       & 119.6         \\
 $R_{cc}, mm$          & 194.5          & 194.5       & 202.0         \\
 $k_c, \%$             & 5.38           & 5.61        & 6.07          \\
 $Z_e, \frac{MOm}{m}$  & 22.54          & 15.57       & 16.61         \\  
 $Q_a$                 & 20170          & 18560       & 21050         \\
\hline
\end{tabular}
\label{2t}
\end{center}
\end{table} \subsection{Thermal-stress analysis}
The thermal-stress analysis for ACS cells has been performed both assuming moderate cooling conditions - the water flow 
velocity $\leq 2 \frac{m}{s}$, input temperature $27 C^o$ and the heat exchange coefficient of $\approx 9500 \frac{W}{m^2 C^o}$. For the initial ($\beta=0.508$) and 
the final ($\beta=0.708$) ACS cell the heat loading corresponding both to $3\%$ and $15\%$ duty factor operation was considered, 
\cite{acsheat},\cite{joshi}. In Fig. 12 the temperature and displacement distributions are shown for $\beta=0.508$ assuming $15\%$ duty 
factor operation. Numerical results are listed in the Table 3.\\
\begin{figure}[htb]
\centering
\epsfig{file=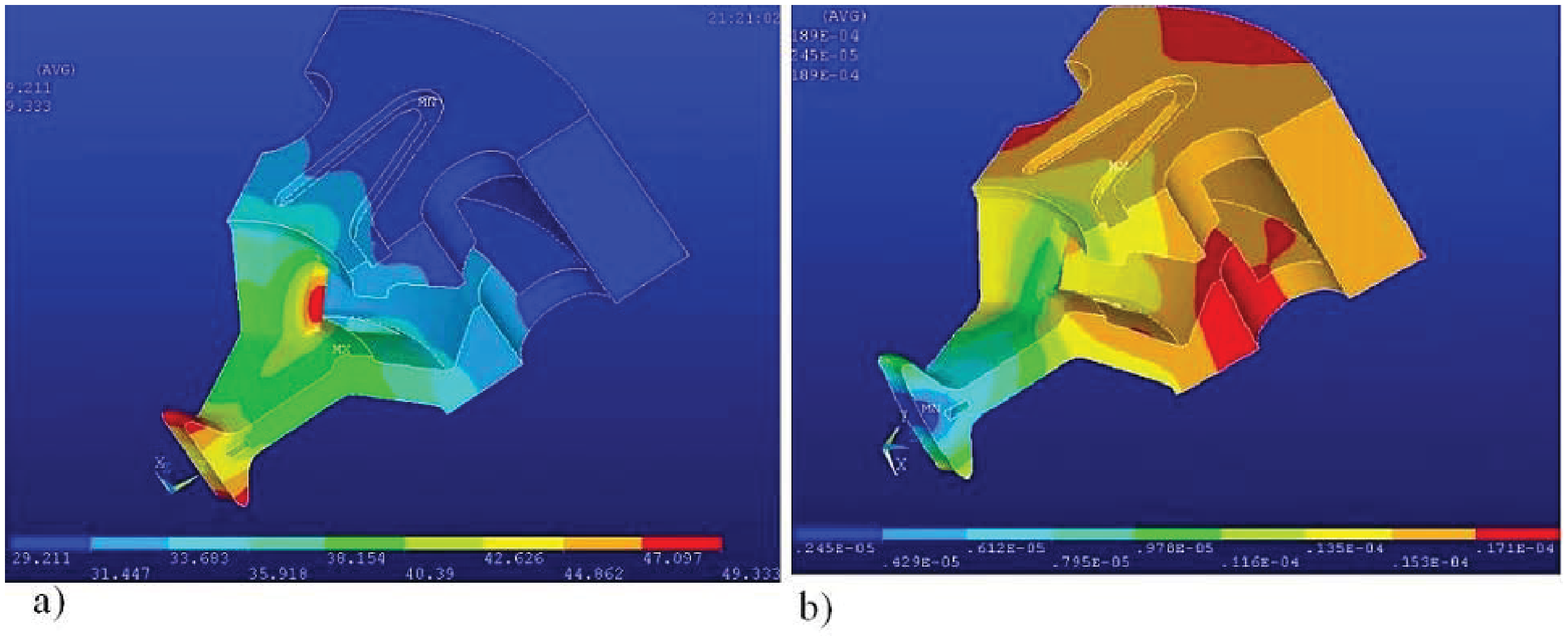, width =145.0mm}
\begin{center}
 Figure 12: The temperature (a) and displacements (b) distributions for $15\%$ duty factor operation, $\beta =0.558$.
\end{center}
\label{12f}
\end{figure}
\begin{table}[htb]   
\begin{center}
\centering{Table 3: The results of structural analysis for ACS cells.}
\begin{tabular}{|l|c|c|c|c|c|c|}
\hline
 $\beta$  &$P_h, \frac{kW}{m}$ & $\delta T_{max}, C^o$ &$\delta f_a, kHz$, & $\delta_c, kHz$ \\
\hline         
 0.5583   & 15.5               & 4.4                   & -42.2             & 2.72            \\
 0.5583   & 77.5               & 22.9                  & -211              & 13.6            \\
\hline
 0.7085   & 12.2               & 4.2                   & -35.6             & -2.6            \\
 0.7085   & 61.0               & 22.1                  & -178              & -13.0           \\
\hline
\end{tabular}
\label{3t}
\end{center}
\end{table} 
As one see from the Fig. 12 and the Table 3, even for very high heat loading $P_h$ the temperature of the drift tube nose $\delta T_{max}$ 
is significant, 
but not drastically. The cooling channels come closer to the drift tube, as possible, and the tube shape results in sufficient heat removal due 
to copper conductivity. It supports the decision about drift tube cone angle in the ACS accelerating cell optimization. As one can see from Fig. 12a, the 
temperature distribution at the surface of accelerating cell (except drift tube nose and near slot region) is rather uniform. The 
shift of accelerating mode frequency $\delta f_a$ is equivalent to the average heating of accelerating cell at $(12-13) C^o$. This value can be 
reduced by increasing of water flow velocity in the safe limits and corresponding increasing of the heat exchange coefficient.  
To compensate this frequency shift for the total ACS module the tuning ability of tuners in the bride cavity is sufficient.\\
Specially should be pointed out the very small in a value and opposite sign for low and high $\beta$, the frequency shift for the 
coupling mode $\delta f_c$, even for $15\%$ duty factor operation. It means, that there will be no stop band width increasing and strong 
deterioration of accelerating field distribution in operation with very high heat loading.\\
The second hot spot in the temperature distribution is at the coupling slot ends. It is common to slot coupled structures and 
takes place due to higher density of RF currents. As compared to another structures, in the current ACS design this effect is reduced 
as possible by the slot shape adjustment and cooling channels choice. But the bottom slot corners are the regions with maximal 
stress value and for $15\%$ duty factor operation the maximal stress value comes close to the yield stress of the copper material. 
It is one of limiting points for possibility of operation even with higher heat loading. 
\subsection{Conclusion for ACS regular cells}
As the result of ACS cells optimization the transverse dimensions of ACS at operating frequency $972 MHz$ are the same as for 
the L-band ACS. And the weight per unit length is even lower, because we withdraw more material from cylinder. It allows 
reliable application of the previously developed fabrication technique, based on brazing with high temperature silver alloys.\\ 
The previously achieved ACS performances are not lost and even improved slightly in $k_c$ value, resulting in improvement of 
uniformity and stability of accelerating field distribution.\\
The mechanical cells design is also sufficiently strong to ensure stability both in construction and during long term stability.\\
In the J-PARC ACS cells neither highest possible $Z_e$ nor $E_s$ values are realized. The goal was to create the counterbalanced 
structure design, when there are no bright achievements in one point at the expense of significant reduction in another parameters, 
but also there are no extra reserve in one point, which allow significant improvement of the total set of parameters.     
\section{Bridge coupler part optimization}
The multi cell Bridge Cavity (BC), developed for L-band ACS module, \cite{acsbrid} consists from similar simple cells, excited in 
$\pi/2$ mode. The schematic sketch of two BC cells together with field distribution, is shown in Fig. 16. For ACS module the 9 cells 
BC is applied. In the excited BC cells the tuners are installed for operating frequency control. Distinguishing from 
coupling coefficient value for regular ACS structure $k_c$, let us denote the coupling coefficient for BC cells as $k_b$. To the ACS
tanks with the regular ACS structure BC is connected with intermediate coupling cells, which have not symmetrical coupling $k_1$ to 
ACS tanks and $k_2$ to BC, $k_2 > k_1$.
\begin{figure}[htb]
\centering
\epsfig{file=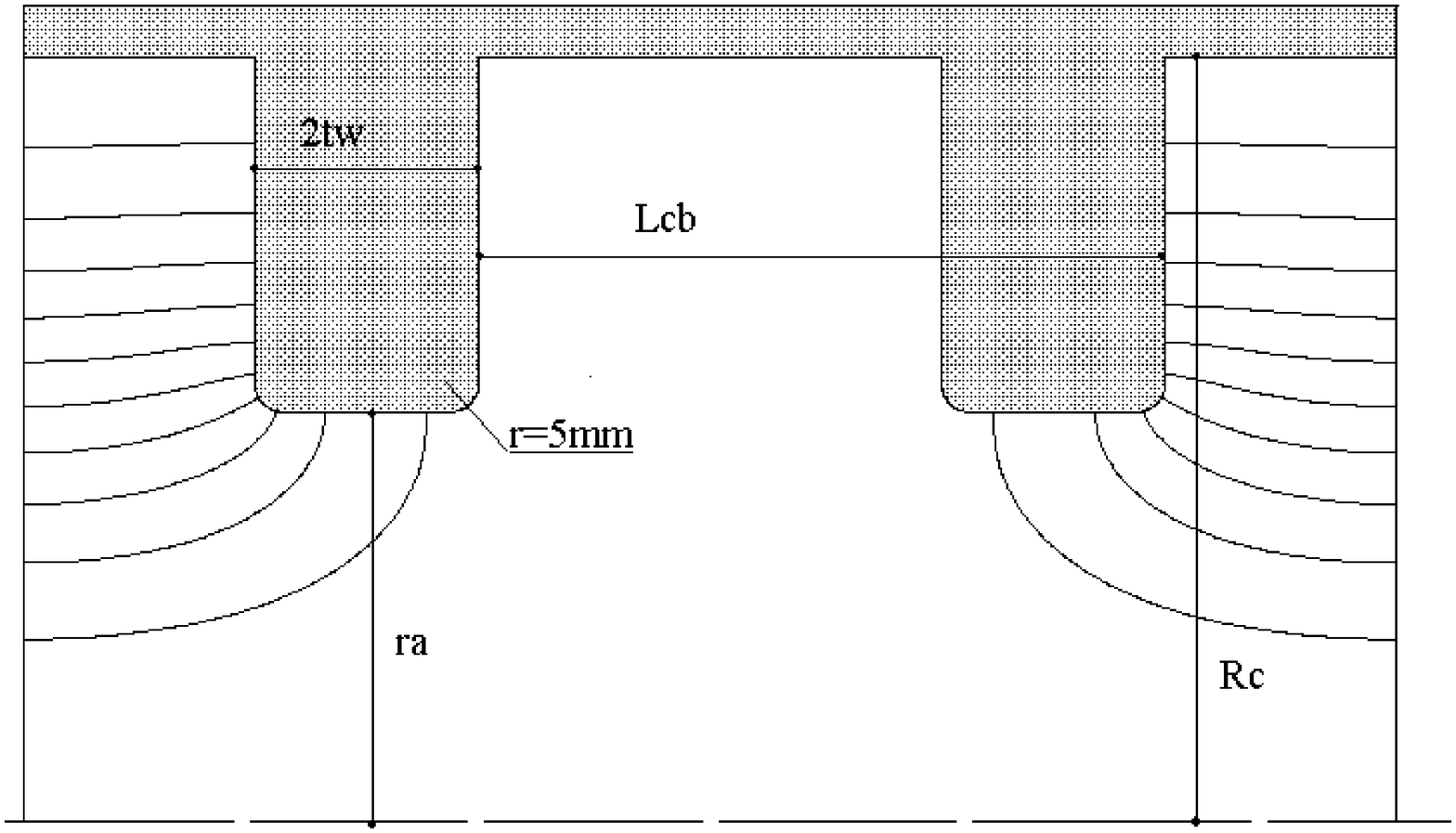, width =95.0mm}
\begin{center}
 Figure 13: Schematic sketch for two regular BC cells with field distribution. 
\end{center}
\label{13f}
\end{figure}
\subsection{The coupling coefficients relations in the ACS module}
The ratio $\frac{k_2}{k_1}$ defines the relative level of the field in excited BC cells with respect to the filed in regular ACS cells.
With $\frac{k_2}{k_1}$ increasing we reduce field in BC cells, reduce RF power, dissipated in BC, but simultaneously reduce tuning rate 
for tuners and increase the value for coupling correction for matching window in RF input cell. For the L-band ACS the ratio $\frac{k_2}{k_1}=2$  
was proven in the high RF power test. No reasons were found in additional consideration to change it and for J-PARC ACS $\frac{k_2}{k_1} =2 $ is adopted 
as the reasonable, safe compromise.\\
\begin{figure}[htb]
\centering
\epsfig{file=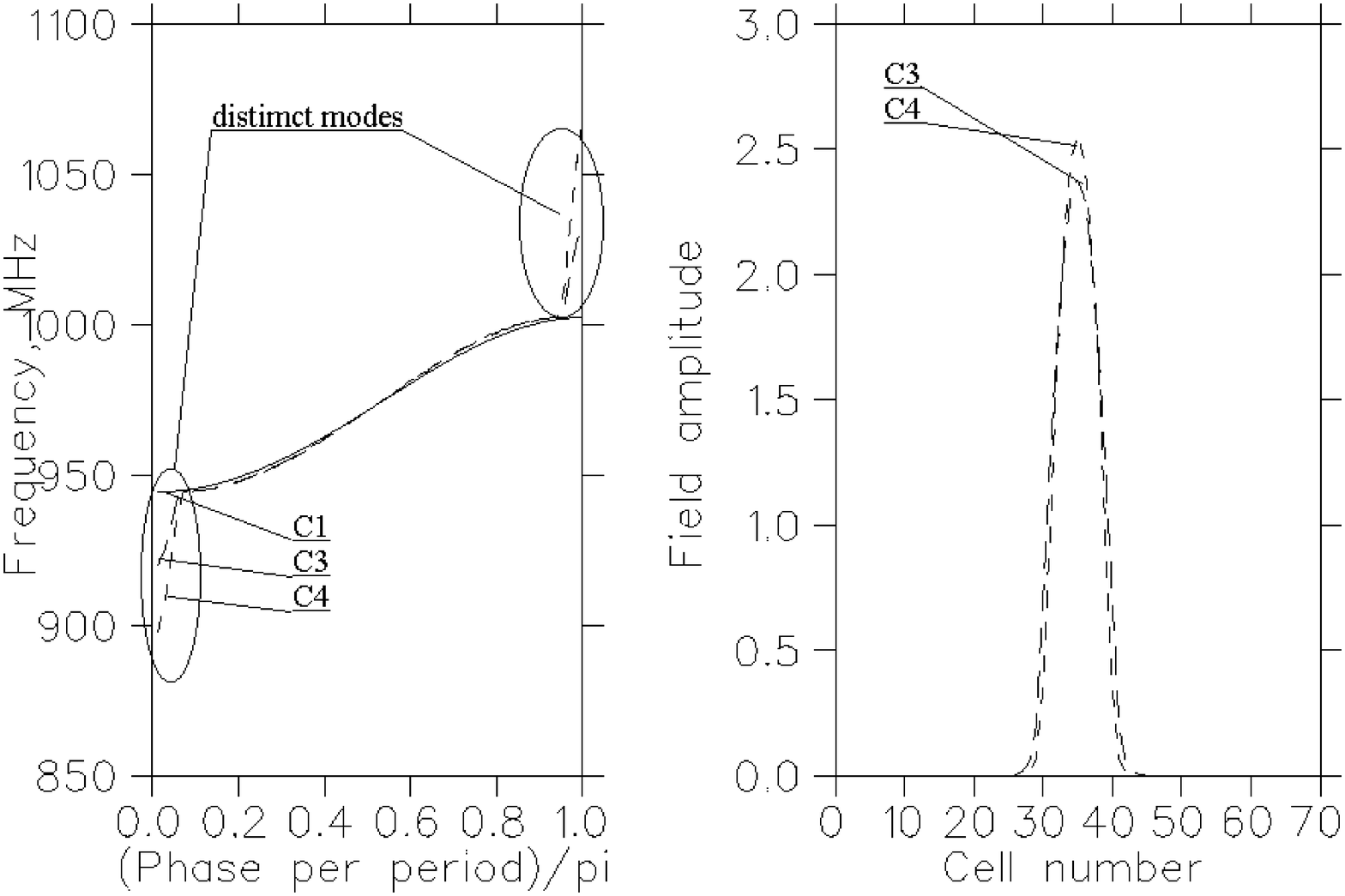, width =115.0mm}
\begin{center}
 Figure 14: Left - the ACS module modes spectrum for different $k_b$ values.
$k_{c}=6\%, k1=6\%, k_2=12\%$, C1) - $k_b=6\%$, C3)- $k_b=12\%$, C4) -
$k_b=18\%$. Right - the amplitude distribution along the ACS cavity
for distinct modes.
\end{center}
\label{14f}
\end{figure}
To avoid frequency spectrum deterioration and field stability reduction, in the ACS module should be fulfilled condition 
$k_1 \geq k_c$. \\
The intermediate coupling cell - this case it is a side coupled cell with respect ACS accelerating cell - was changed from circular to more complicated shape to fit 
with modified ACS accelerating cell. The required $k_1$ value is achieved by increased slot opening to $49.8^o$. The value of another 
coupling coefficient $k_2$ depends on the distance between BC axis and intermediate cell nose axis, which defines the insertion 
of intermediate cell into the first BC cell.\\    
The coupling coefficient $k_b$ for BC cells is not fixed so rigidly. As one can see from BC cell sketch, Fig. 13, for the fixed length 
of the BC cell $L_{cb}$ we have three variables - the cell radius $R_c$, the aperture radius $r_a$ and the disk thickness $t_w$. But the 
rigid requirement is just one - operating frequency $972MHz$. Additionally we have to provide sufficient coupling coefficient, frequency 
separation with the nearest $TE_{11n}$ passband and BC vacuum conductivity. In the BC cavity for L-band ACS the concept $t_w=const$ 
was suggested, resulting in the high $k_b \approx 18\%$ value and sufficient separation with $TE_{11n}$ passband.\\
For the case $k_b \gg k_c$ in the modes spectrum of ACS module arise distinct modes, which are placed enough far outside from ACS 
passband, Fig. 14 (left). The field for such modes is concentrated in BC cells and do not penetrate in ACS tanks, Fig. 14 (right). Such 
modes do not improve the field stability in ACS module, which is defined by coupling coefficient of regular ACS cells. From this 
point of view, very big $k_b$ value is an extra reserve and do not improve significantly te ACS module parameters.\\
For mass production looks attractive to fix the cell radius $R_c$ for all BC's in the system. Additionally, the distance from the 
BC axis to the beam axis will be constant over entire ACS system, for more standard in connections.\\
In the BC's for J-PARC ACS the concept of 'constant cell volume' is realized. If we need in adjustment for two dimensions (for 
$L_{cb}$ increasing with $\beta$) , $t_w$ and $r_a$, let us connect the disk thickness as $L_{cb}-t_w=80.0 mm$. This case for low $\beta$ 
we have sufficient for rigidity disk thickness $2t_w=12.72 mm$ and $k_b \approx 12\%$, (depending on $R_c$ choice). With this condition the volume 
of the field in BC cell remains approximately the same for all $\beta$. It results in 'nearly the same' tuning rate for tuners 
over entire ACS system and in 'nearly the same' $k_2$ value in BC coupling with intermediate cell. Such solution significantly 
decreases the possible spread in tuning rate and transverse dimensions between different module's in ACS system. The frequency 
separation with the nearest $TE_{11n}$ passband in BC cells even improves - the cells becomes shorter for high $\beta$ region.\\
The BC's cell radius was chosen to have $ 14.9\% > k_b > 7.8\%$ and frequency separation between ACS passband and  $TE_{11n}$ passband 
is $> 60 MHz$ over entire 
ACS system. Because the aperture radius $r_a$ rises with $\beta$, simultaneously with $2t_w$ rise, the vacuum conductivity of 
BC is not deteriorated. For BC's cells in ACS module is the fixed value $R_c=132.2 mm$.\\       
\begin{figure}[htb]
\centering
\epsfig{file=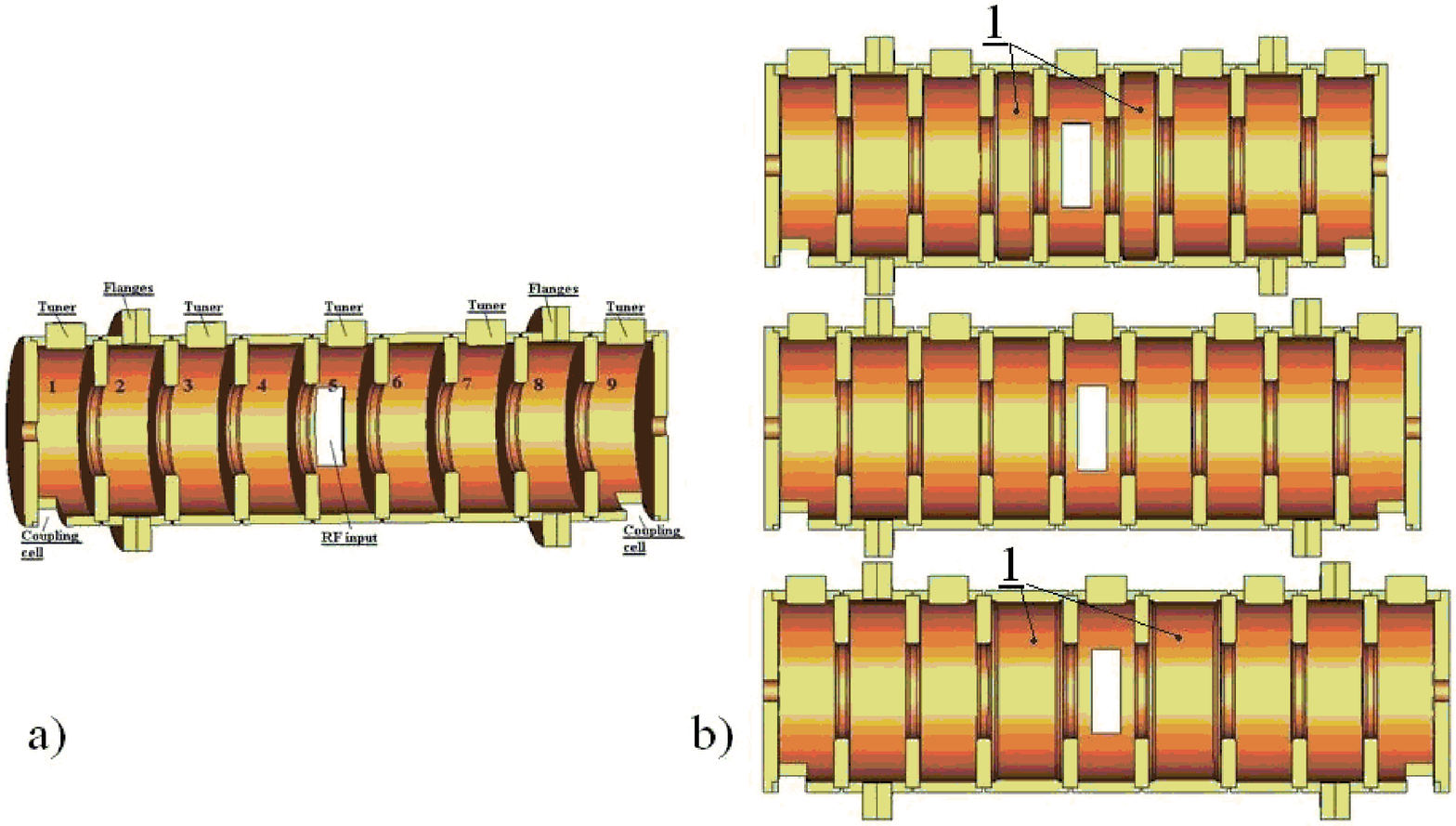, width =135.0mm}
\begin{center}
 Figure 15: Schematic sketches bridge cavity design (a) and bridge cavities with variable coupling cells (b).
\end{center}
\label{15f}
\end{figure}
The schematic sketch for BC cavity design, adopted for J-PARC ACS module, is shown in Fig. 15a.\\ 
In optimization for mass production, due to the request of productor, the additional possibility was considered. 
Following to $\beta$ increasing, the total BC length should increases. It can be realized by 
changing the length of all cells in BC simultaneously. But we can do it by changing length just for several cells and keeping the 
length of another cells as constant for different BC's cavities. This case all BC cavities should be reasonably separated in several 
groups. In each group BC have the cells with the constant, inside the group, length, and one or two cells with variable length, specific 
for each BC. Subdivision in groups is required, because the possibility in cell length change is limited by RF parameters.  
For cell length decreasing we have increasing of the cell frequencies for operating $TM_{010}$ mode, both for the cell under change and 
for neighbor cells. For cell length increasing we have decreasing of $TE_{11n}$ modes frequencies and also decreasing for 
operating mode frequencies, both for the cell and for neighbors.\\
The most convenient choice for cells with variable length is two not excited cells near the middle BC cell with RF input. This case the own 
frequency for the variable cell can be adjusted by cell radius $R_{c'}$, which becomes specific for each variable cell. The distortion 
of frequency for neighbor cells, which are excited and equipped with tuners, is of $\approx 0.5 MHz$ and can be corrected by tuners.
The variation of $k_b$ value is not significant $\delta k_b/k_b < 7\%$ and passband separation with $TE_{11n}$ modes is $>50 MHz$.
Such solution leads to decreasing in the set of BC dimensions in mass production, but requires the individual tuning of variable cells and 
individual tuners adjustment for neighbor cells in ACS operation. This solution was not accepted for ACS. 
\subsection{The shape of matching window}
The vicinity of matching window between driving waveguide and input cell in the bridge cavity was modified as shown in Fig. 16. 
The initial window configuration is shown in Fig. 16a and has not the best cooling conditions. Together with increased density of RF 
current near edges, it can result in the temperature rise.\\
\begin{figure}[htb]
\centering
\epsfig{file=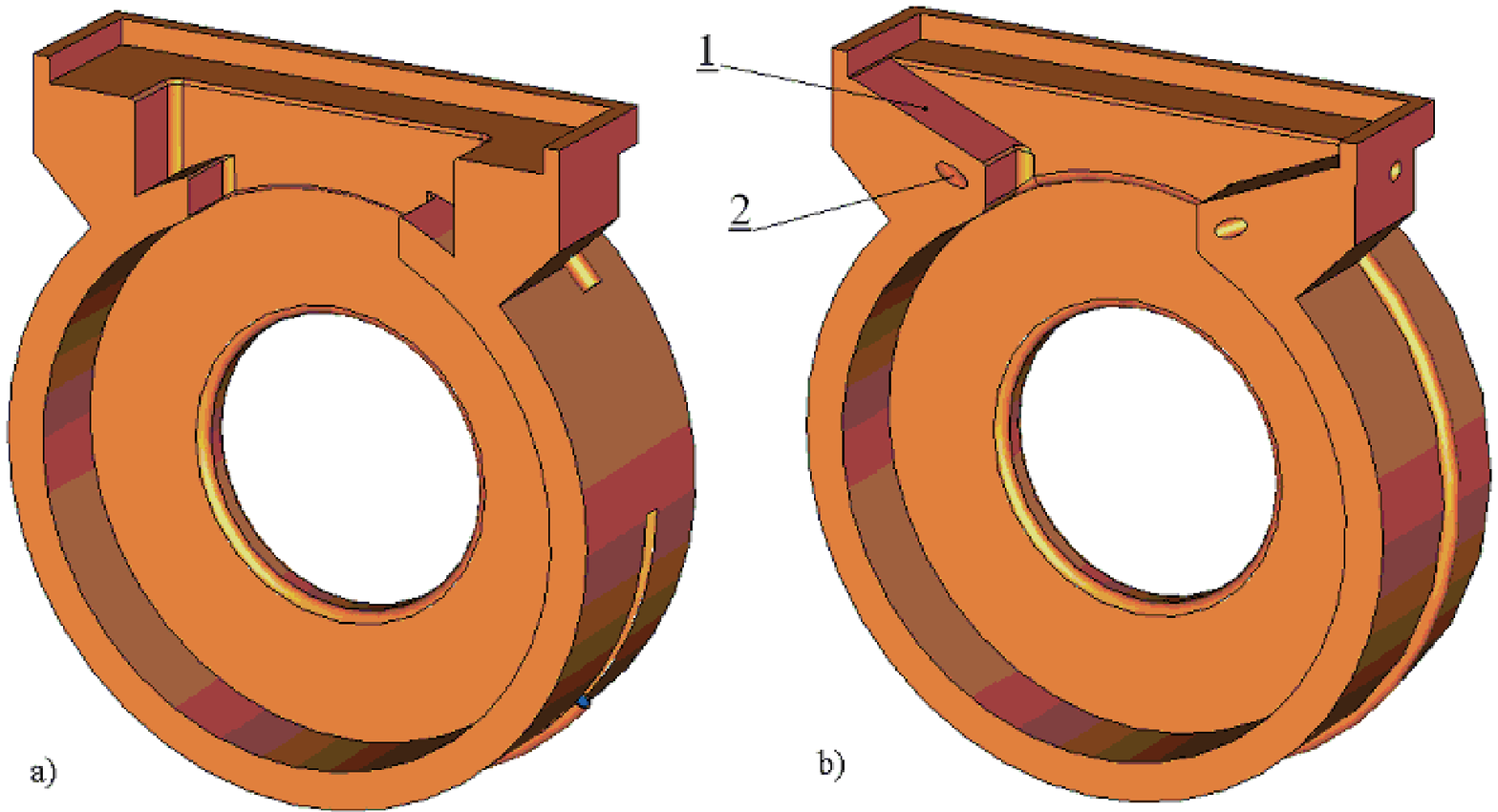, width =115.0mm}
\begin{center}
 Figure 16: Modification of the matching window. 
\end{center}
\label{16f}
\end{figure}
The idea of tapering near window, similar to tapering in ACS coupling cells, was applied, Fig. 16b. The width of the window (in the beam 
axis direction for ACS module) is limited for low $\beta $ by the length of input cell in bridge cavity. The tapering was applied for the 
window length. Additional tapering in perpendicular direction is not effective, because leads to the coupling reduction. 
With the tapering near modified window the same coupling is obtained with shorter window, resulting in reduction of RF current density, 
which is reduced additionally by edges rounding. In this design the space is sufficient to place V-like cooling channel close to window,
providing better conditions for matching window cooling.  
\section{Comparison with SCS in parameters}
After all modifications, the ACS cell were compared in parameters with SCS cells. SCS structure is mostly distributed in hadron 
linacs and can be considered as the reference one. But SCS is applied at the frequency $f_{op}=805 MHz$, sometimes with special 
requirements, such as high accelerating gradient in the FNAL linac. We have to compare the structures at other equal conditions - 
frequency, aperture radius and 
so on. For this purpose we considered for the SCS the ACS-like accelerating cell in the dimensions of drift tube and web thickness - it were chosen 
taking into account J-PARC linac requirements at the frequency $972 MHz$. But for SCS we accept the outer cell rounding, Fig. 2b, Fig. 17a, even 
supposing an initial preference in $Z_e$ for SCS. The conical part in the shape of SCS cell is not required. For coupling cells formation 
was accepted concept, realized for SNS coupled cell linac \cite{scssns}. The direct scaling of SCS coupling slots from SNS design 
results in $k_c=4.32\%$, compared to $k_c=5.86\%$ in ACS for $\beta=0.556$. Structures should be considered at the same $k_c$ value.\\ 
\begin{figure}[htb]
\centering
\epsfig{file=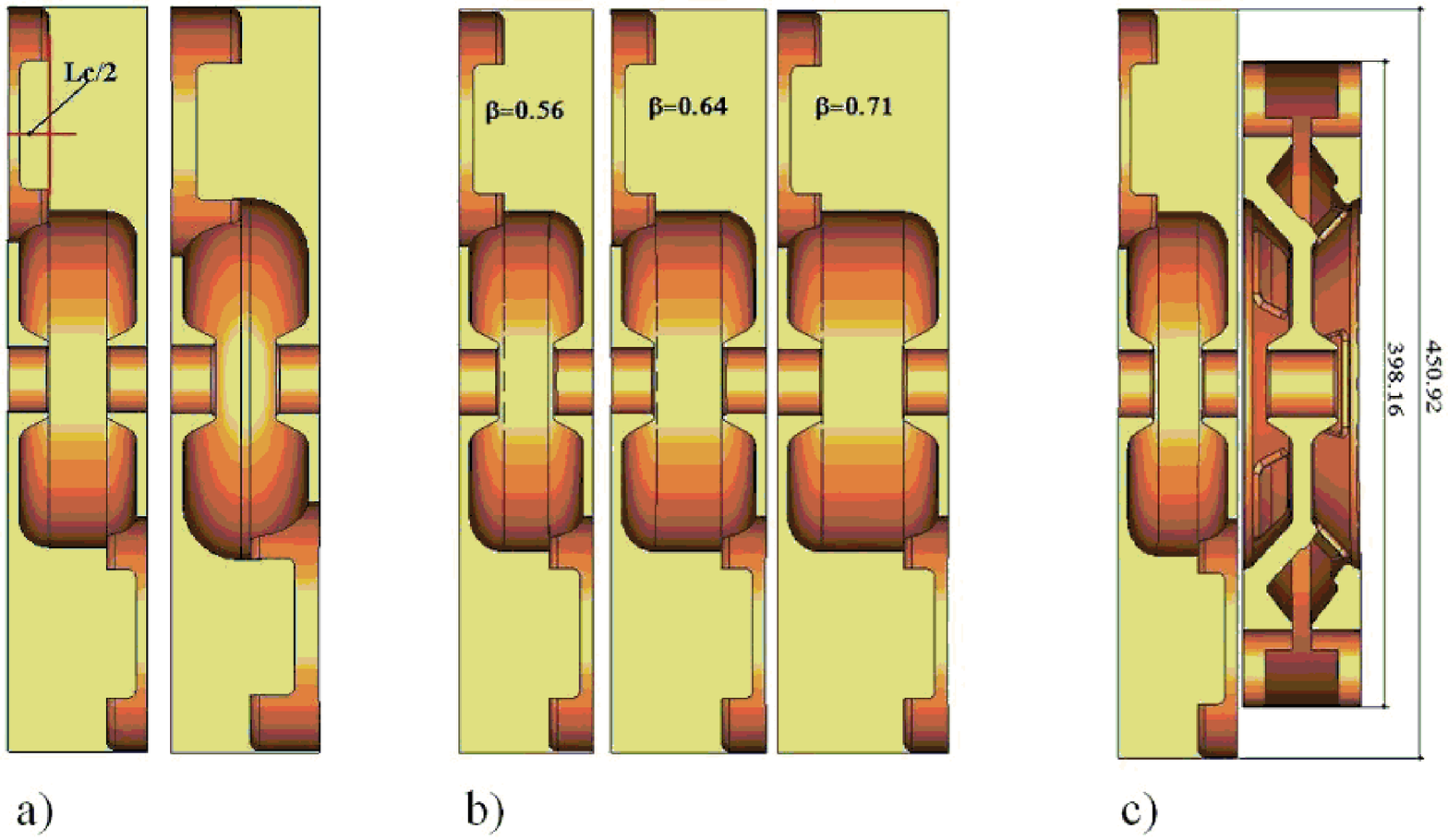, width =115.0mm}
\begin{center}
 Figure 17: ACS - SCS comparison. a) - coupling coefficient increasing for SCS, b) obtained SCS configurations, c) SCS and ACS 
comparison for $f_0=972 MHz$.
\end{center}
\label{17f}
\end{figure}
In SCS the coupling coefficient value $k_c$ can be increased by coupling cell length $L_c$ increasing, together with more deep coupling 
cell insertion into accelerating one, see Fig. 1a. It leads to strong increasing in coupling slot dimensions, rise in direct coupling 
coefficient $k_c$ value. But simultaneously much faster rises the coefficient of neighbor coupling $k_{aa}$, resulting in the distortion 
of dispersion curve. The limiting case of coupling cell insertion is shown in Fig. 17a (right) and is limited by crossing with 'drift 
tube' in coupling cell. Further insertion of coupling cell into accelerating cells does not results in $k_c$ increasing.  With 
the single SCS slot against four ACS slots we have no freedom in $k_c$ increasing and should go to not comfortable mechanical design of 
the slot region. Anyhow, adjusting $k_c =5.57\%$ for $\beta =0.556$, we apply, according procedure of the SNS coupled cell linac, the same 
coupling cell to the total $\beta$ range. Examples of the obtained SCS are illustrated in Fig. 17b.
Results of comparative simulations are listed in the Table 4.\\
\begin{table}[htb]   
\begin{center}
\centering{Table 4: Comparison for SCS and ACS options at the J-PARC linac conditions.}
\begin{tabular}{|l|c|c|c|c|c|c|}
\hline
 $\beta$                      & 0.56        &  0.64        & 0.71 \\
\hline          
$\frac{k_{cSCS}}{k_{cACS}}$   & 5.57/5.86   &  5.18/5.58   &  4.90/5.32            \\
\hline
$\frac{Z_{eSCS}}{Z_{eACS}}$   & 0.970       & 0.993        &  1.005                \\
\hline
$\frac{k_{aaSCS}}{k_{aaACS}}$ & 0.75/0.01   &  0.66/0.01   &  0.60/0.01            \\
\hline
$\frac{k_{ccSCS}}{k_{ccACS}}$ & 0.01/0.63   & 0.01/0.31    &   0.01/0.2            \\
\hline
\end{tabular}
\label{4t}
\end{center}
\end{table} 
As one can see from the Table 4, even for lower $k_c$ value, for the same conditions, SCS has no preference in RF efficiency.
And initial preference due to rounded cell shape in cancelled by not comfortable connections of the cells. In the shape distortions for 
dispersion curve, the structures are nearly the same - there is neighbor coupling between ACS coupling cells and similar coupling between SCS 
accelerating cells. \\
More comparison is in results of thermal-structural analysis. For SNS the operational duty factor is of $6\%$. The results of 
such SCS thermal analysis are described in \cite{scsslotheat}. Comparing results for SCS and ACS, one will see less uniform temperature distribution for SCS 
and higher stress values near slot, even for adopted coupling cell for SCS design, which leads to lower $k_c$ value for J-PARC conditions.\\
The J-PARC ACS accelerating module has much more reserve for operation with the enlarged duty factor value.
\section{Summary}
Proposals for the Annular Coupled Structure modification to improve ACS 
advantages for 972 MHz option are described in this report. Mainly these 
improvements are directed to simplify mass production and mass tuning process.
But each modification should not deteriorate (only improve) ACS parameters 
achieved before.
For this purpose all modifications were strongly investigated with different 
methods, available in the design of accelerating structures, discussed 
and revised for possible sequences. Only if in the reference ACS design 
there was extra reserve in some parameters, these reserves were decreased 
to conservative, proven in another structures, values, leading to improvements
in another parameters.\\
Each modification separately is not a big achievement. But collected and 
adjusted all together, these modifications have enough significant 
effect - proposed ACS design doesn't lose to reference design in rf parameters
but has advantage of smaller transverse dimensions and strongly reduced 
weight, allowing application of fabrication technique similar to L- band 
ACS option.\\
During JHP research program very robust concept of the accelerating tank 
was developed and tested - reasonably effective accelerating structure with 
very effective cooling circuit, added with bridge coupling cavities, equipped 
with fast movable tuners. Together with latest improvements in ACS design 
and follow the concept of the total cavity, ACS looks now as the mostly 
promising cost-effective solution for normal conducting accelerating structure
for high intensity proton linac with high duty factor operation. The present 
ACS 972 MHz design has the possibility to operate with the average heat loading 
up to $80 kW/m$ during $15\%$ duty factor
 linac operation.    
\section{Acknowledgments}   
This report presents one part of the 972 MHz ACS development, mainly results 
of RF parameters study, which was under way in KEK (and later JAEA) ACS group. All results 
were discussed in ACS meetings - nice trial field to protect 
and improve proposals in the atmosphere of creative research, 
large experience and high responsibility for final result - high intensity 
linac.\\
The author warmly thanks all participants of ACS meetings - Y. Yamazaki, H. Ao,
M. Ikegami, F. Naito, T. Kato, N. Hayashizaki, S.C. Joshi, A.Ueno, for providing creative 
atmosphere of the joint work, discussions, recommendations, reasonable 
objections and all time - warm human relations.    

\end{document}